\documentstyle[12pt,amsfonts]{article}
\newcommand{\svec}[1]{ \stackrel{\rightarrow}{#1} }

\newcommand{\ehat}{ \hat U_{\epsilon} }

\newcommand{\define}{ \stackrel{\triangle}{=} }

\def\be{\begin{equation}}
\def\ee{\end{equation}}
\def\ba{\begin{array}}
\def\ea{\end{array}}

\def\d4{{\rm d}^4}

\begin{document}
\vskip -1.0cm
\title{\bf Coupling Between the Spin and Gravitational Field and
            the Equation of Motion of the Spin}
\author{ {Ning WU}\thanks{email address: wuning@mail.ihep.ac.cn}
\\
\\
{\small Institute of High Energy Physics, P.O.Box 918-1,
the Chinese Academy of Science,} \\
{\small Beijing 100039, P.R.China}
}
\maketitle


\begin{abstract}
In general relativity, the equation of motion of the spin is given
by the equation of parallel transport, which is a result of the space-time
geometry. Any result of the  space-time geometry can not be directly
applied to gauge theory of gravity. In gauge theory of gravity,
based on the viewpoint of the coupling between the spin and gravitational
field, an equation of motion of the spin is deduced. In the post
Newtonian  approximation, it is proved that this equation gives out the same
result as that of the equation of parallel transport. So, in the post Newtonian
approximation, gauge theory of gravity gives out the same prediction
on the precession of orbiting gyroscope as that of general relativity.
\end{abstract}



\newpage

\Roman{section}

\section{Introduction}

It is known that any theory on gravitational interactions
should first pass classical tests of gravity. General Relativity(GR)
has passed several classical tests\cite{01,02}, including
the deflection of light by the sun\cite{01,021},
the precession of the perihelia of the orbits of the
inner planets\cite{01,022}, and the time delay of radar echoes passing
the sun\cite{023,024}. Another new test, the precession of orbiting
gyroscopes\cite{025,026,027}, is going on the way. It is reported that,
with an accuracy of about 20\%, the observed precession of the
orbiting gyroscopes is consistent with the predictions of GR\cite{028,028a}.
Basis of the calculation of these classical
tests in GR is Schwarzschild solution, geodesic
equation, and the equation of parallel transport\cite{03,04}.
\\

The precession of orbiting gyroscopes is a gravitomagnetic effect.
The classical effects of gravitomagnetism were studied for more
than one hundred years.
The close analogy between Newton's gravitation law and Coulomb's law
of electricity led to the birth of the concept of gravitomagnetism
in the nineteenth century\cite{m01,m02,m03,m04}. Later, in the
framework of GR, gravitomagnetism was extensively
explored\cite{m05,m06,m07}. Some recently reviews
on gravitomagnetism from different viewpoints
can be found in literatures \cite{m08,m09,m10}.
\\

Quantum Gauge Theory of Gravity(QGTG) is first proposed in
2001\cite{05,06,07,08}. The motivation to propose QGTG is try
to unify general relativity with quantum theory in the framework
of gauge field theory. This goal does not reached until 2003,
when Quantum Gauge General Relativity(QGGR)
is proposed in the framework of QGTG\cite{09,10}.
QGGR is a perturbatively renormalizable
quantum theory, so based on it, quantum effects of
gravity\cite{11,11a,12,13} and gravitational interactions of some basic
quantum fields \cite{14,15} can be explored. Unification of fundamental
interactions including gravity can be fulfilled in a simple and
beautiful way\cite{16,17,18}. If we use the mass
generation mechanism reported
in the literatures \cite{19,20}, we can propose a new
theory on gravity which contains massive graviton and
the introduction of massive graviton does not affect
the strict local gravitational gauge symmetry of the
Action and does not affect the traditional long-range
gravitational force\cite{21}. The existence of massive graviton
will help us to understand the possible origin of dark matter.
In QGGR, the field equation of gravitational gauge field is just
Einstein's field equation, so in classical level, we can set up
its geometrical formulation\cite{21a}, and QGGR returns to
Einstein's general relativity in classical level.  In QGGR, the
field equation of gravitational gauge field is the same as
Einstein's field equation in GR, so two equations
have the same solutions, though mathematical expressions of the
two equations are completely different\cite{21b}. For three
classical tests, the deflection of light by the sun,
the precession of the perihelia of the orbits of the
inner planets, and the time delay of radar echoes passing
the sun, QGGR gives out the same predictions as those of
GR\cite{21b}.
\\

In this paper, the equation of motion of the spin is discussed
in the framework of QGGR. In QGGR,  gravity is treated as a kind
of physical interactions, which is transmitted by graviton. So in
QGGR, the equation of parallel transport can not be directly used,
for it is a result of space-time geometry.  Based on the coupling
between the spin and gravitational field, the equation
of motion of the spin can be obtained. Then we prove that this
equation and the equation of parallel transport give out the same
results in the post Newtonian  approximation.
 \\

\section{Gravitomagnetic and Gravitoelectric Field}
\setcounter{equation}{0}

For the sake of integrity, we give a simple
introduction to QGGR and introduce
some notations which is used in this paper. Details on
QGGR can be found in
literatures \cite{05,06,07,08,09,10}.
In gauge theory of gravity, the most fundamental
quantity is gravitational gauge field $C_{\mu}(x)$,
which is the gauge potential corresponding to gravitational
gauge symmetry. Gauge field $C_{\mu}(x)$ is a vector in
the corresponding Lie algebra, which, for the sake
of convenience, will be called gravitational Lie algebra.
So $C_{\mu}(x)$ can be expanded as
\be \label{2.1}
C_{\mu}(x) = C_{\mu}^{\alpha} (x) \hat{P}_{\alpha},
~~~~~~(\mu, \alpha = 0,1,2,3)
\ee
where $C_{\mu}^{\alpha}(x)$ is the component field and
$\hat{P}_{\alpha} = -i \frac{\partial}{\partial x^{\alpha}}$
is the  generator of global gravitational gauge group.
The gravitational gauge covariant derivative is given by
\be \label{2.3}
D_{\mu} = \partial_{\mu} - i g C_{\mu} (x)
= G_{\mu}^{\alpha} \partial_{\alpha},
\ee
where $g$ is the gravitational coupling constant and matrix
$G = (G_{\mu}^{\alpha}) = ( \delta_{\mu}^{\alpha} - g C_{\mu}^{\alpha} )$.
Its inverse matrix is
$G^{-1} = \frac{1}{I - gC} = (G^{-1 \mu}_{\alpha})$.
Using matrix $G$ and $G^{-1}$, we can define two important
composite operators
\be \label{2.6}
g^{\alpha \beta} = \eta^{\mu \nu}
G^{\alpha}_{\mu} G^{\beta}_{\nu},
\ee
\be \label{2.7}
g_{\alpha \beta} = \eta_{\mu \nu}
G_{\alpha}^{-1 \mu} G_{\beta}^{-1 \nu},
\ee
which are widely used in QGGR. In QGGR,
space-time is always flat and space-time metric
is always Minkowski metric, so $g^{\alpha\beta}$ and $g_{\alpha\beta}$
are no longer space-time metric. They are only two composite operators
which  consist of gravitational gauge field. \\

The  field strength of gravitational gauge field is defined by
\be \label{2.8}
F_{\mu\nu} (x) \define \frac{1}{-ig} \lbrack D_{\mu}~~,~~D_{\nu} \rbrack =
F_{\mu\nu}^{\alpha}(x) \cdot \hat{P}_{\alpha}
\ee
where
\be \label{2.12}
F_{\mu\nu}^{\alpha} =
G_{\mu}^{\beta} \partial_{\beta} C_{\nu}^{\alpha}
-G_{\nu}^{\beta} \partial_{\beta} C_{\mu}^{\alpha}.
\ee
Define
\be \label{2.12a}
F^{\alpha}_{ij}= - \varepsilon_{ijk} B^{\alpha}_{k}
~~,~~
F^{\alpha}_{ 0i} = E^{\alpha}_{i}.
\ee
Then field strength of gravitational gauge field can be expressed
as
\be
F^{\alpha} = \left \lbrace
\ba{cccc}
0 & E_1^{\alpha} & E_2^{\alpha} & E_3^{\alpha}  \\
- E_1^{\alpha} & 0 & -B_3^{\alpha}  &  B_2^{\alpha}  \\
- E_2^{\alpha} & B_3^{\alpha} & 0  & - B_1^{\alpha}  \\
- E_3^{\alpha} & -B_2^{\alpha}  &  B_1^{\alpha} & 0
\ea
\right \rbrace .
\ee
This form is quite similar to that of field strength in
electrodynamics, but with an extra group index $\alpha$.
The component $E_i^{\alpha}$ of field strength is called
gravitoelectric field, and $B_i^{\alpha}$ is called
gravitomagnetic field. Traditional Newtonian gravity
is transmitted by gravitoelectric field
$E_i^{0}$, and the gravitational
Lorentz force is transmitted by gravitomagnetic field
$B_i^{\alpha}$\cite{23}. Effects of gravitomagnetic
field in astrophysical processes maybe observable\cite{13}.
The equation of motion of the
spin discussed in this paper is dominated by the coupling
between the spin of the particle and gravitomagnetic field.
The $\alpha=0$ components $B_i^0$ and $E_i^0$
respectively correspond to the gravitomagnetic field
and gravitoelectric  field defined in
literature \cite{m05,m06,m07,m08,m09,m10}.
\\

\section{The Equation of Motion of the Spin in Gravitational Field}
\setcounter{equation}{0}

First, we need to determine the coupling term between the spin of
a particle and gravitomagnetic field. In electromagnetic interactions,
the coupling energy between the spin $\svec{J}$ of a particle and
magnetic field $\svec{B}_e$ is
\be \label{3.1}
-  \frac{q}{m} \svec{J} \cdot \svec{B}_e,
\ee
where $q$ is the electric charge of the particle, and $m$ is its
mass. We can conjecture that the coupling between the spin $\svec{J}$
and gravitomagnetic field $\svec{B}^{\alpha}$ should be proportional
to $\svec{J} \cdot \svec{B}^{\alpha}$. This is true. In literature
\cite{15,12}, in the non-relativistic limit and weak gravitational
field approximation, we have deduced that the coupling
between the spin of a Dirac particle and gravitomagnetic field
$\svec{B}^{\alpha}$ is
\be \label{3.2}
-  \frac{g}{2} \svec{\sigma} \cdot \svec{B}^0,
\ee
where $\sigma_i$ is the Pauli matrix. Because the spin operator of
the Dirac particle is $\frac{\sigma_i}{2}$, for a general particle
with spin $\svec{S}$, we can generate the  equation (\ref{3.2}) to the
following form
\be \label{3.3}
\Delta H = -  g \svec{S} \cdot \svec{B}^0
= - g S B^0 \cos \theta,
\ee
where $\Delta H$ is the energy shift caused by this coupling, and
$\theta$ is the angle between directions of spin and gravitomagnetic
field. Then the magnitude of the torque acting on the spin is
\be \label{3.4}
L = - \frac{\partial \Delta H}{\partial \theta}
= - g S B^0 \sin \theta.
\ee
Its direction along the direction of enlarging the $\theta$ angle.
So, the torque vector $\svec{L}$ should be
\be \label{3.5}
\svec{L} = g \svec{S} \times \svec{B}^0.
\ee
From definition (\ref{2.12a}) of gravitomagnetic field, we have
\be \label{3.6}
L_i = -g F_{ij}^0 S_i.
\ee
The above relation gives out the leading contribution of the torque
caused by the gravitomagnetic interaction of the spin, which is in
a non-relativistic form. Indeed, it is obtained in the non-relativistic
limit and  weak gravitational field approximation. According to
the gauge principle and the principle of relativity, any fundamental
equation of motion should be Lorentz covariant and gauge covariant.
Denote the final required torque as $L_{\mu}$. Then under Lorentz
transformations, it should be transformed as
\be \label{3.7}
L_{\mu} \to L^{\prime}_{\mu} = \Lambda_{\mu}^{~\nu} L_{\nu},
\ee
and under gravitational gauge transformations, it should be transformd as
\be \label{3.8}
L_{\mu} \to L^{\prime}_{\mu} = (\ehat L_{\nu}),
\ee
and it should return to equation (\ref{3.6}) in non-relativistic
limit and weak field approximation. We can prove that the
following torque satisfies all above requirements
\be \label{3.9}
L_{\mu} = g \eta^{\lambda \nu} g_{\alpha \beta}
F_{\mu\nu}^{\alpha} {\mathbb S}_{\lambda} U^{\beta},
\ee
where ${\mathbb S}_{\mu}$ is the gravitogauge canonical spin, and
$U^{\beta}$ is the velocity of the particle.
Therefore, the equation of motion of the spin in gravitational
field is
\be \label{3.10}
\frac{{\rm d} {\mathbb S}_{\mu}}{{\rm d} \tau}
= g \eta^{\lambda \nu} g_{\alpha \beta}
F_{\mu\nu}^{\alpha} {\mathbb S}_{\lambda} U^{\beta}.
\ee
\\

The relation between the gravitogauge canonical spin ${\mathbb S}_{\mu}$
and its ordinary spin four-vector $S_{\alpha}$ is
\be \label{3.11}
{\mathbb S}_{\mu} = G_{\mu}^{\alpha} S_{\alpha}.
\ee
Therefore, we have
\be \label{3.12}
G^{-1 \mu}_{\alpha}
\frac{{\rm d} {\mathbb S}_{\mu}}{{\rm d } \tau}
= \frac{{\rm d} S_{\alpha}}{{\rm d} \tau}
+ G^{-1 \mu}_{\alpha}
(\partial_{\gamma} G_{\mu}^{\beta}) S_{\beta} U^{\gamma}.
\ee
Then equation (\ref{3.10}) can be changed into
\be \label{3.13}
\frac{{\rm d} { S}_{\alpha}}{{\rm d} \tau}
= G_{\mu}^{\beta}
( \partial_{\gamma} G^{-1 \mu}_{\alpha})
S_{\beta} U^{\gamma}
+g \eta^{\lambda \nu} g_{ \beta \gamma}
G^{-1 \mu}_{\alpha} G_{\lambda}^{\delta}
F_{\mu\nu}^{\beta} { S}_{\delta} U^{\gamma}.
\ee
This is the equation of motion of the spin $S_{\alpha}$.
\\

\section{Post Newtonian Approximation}
\setcounter{equation}{0}

Now, let's discuss the post Newtonian approximation of the
equation (\ref{3.10}). In post Newtonian approximation, we
have
\be \label{4.1}
\left \lbrace
\ba{rcl}
g_{00} & = & \eta_{00} - 2 \phi + o(\bar v^4) \\
&&\\
g_{ij} & = & \eta_{ij} - 2 \delta_{ij} \phi + o(\bar v^4)\\
&&\\
g_{i0 }&=& g_{0i } = \zeta_i + o(\bar v^5)\\
&&\\
U^0 &=& 1 + o(\bar v^2)\\
&&\\
U^i &=& v^i + o(\bar v^3)\\
&&\\
{\mathbb S}_i &=& S_i + o(\bar v^2 S) \\
&&\\
{\mathbb S}_0 &=& - \svec{v} \cdot \svec{S} + 0(\bar v^3 S)
\ea
\right . ,
\ee
where $\phi$ and $\zeta_i$ are given by the following relations
\be \label{4.2}
\phi = - \frac{G M}{r},
\ee
\be \label{4.3}
\svec{\zeta} = \frac{2 G}{r^3}
(\svec{x} \times \svec{J}).
\ee
In above relations, $M$ and $\svec{J}$ are mass and spin angular
momentum of the source. In post Newtonian approximation, we have
\be \label{4.4}
\frac{{\rm d} {\mathbb S}_i}{{\rm d} \tau}
= \frac{{\rm d} S_i}{{\rm d} t}
+ S_i \frac{\partial \phi}{\partial t}
+ S_i (\svec{v} \cdot \nabla \phi)
+ o(\bar v^4),
\ee
and
\be \label{4.5}
g \eta^{\lambda \nu} g_{\alpha \beta}
F_{ i \nu}^{\alpha} {\mathbb S}_{\lambda} U^{\beta}
=  -2 (\partial_i \phi) (\svec{v} \cdot \svec{S})
+ \frac{1}{2} (\partial_i \zeta_j) S_j
- \frac{1}{2} (\svec{S} \cdot \nabla) \zeta_i
+  v_i (\svec{S} \cdot \nabla \phi)
+ o(\bar v^4).
\ee
In equation (\ref{3.10}), set $\nu=i$ and apply the above results, we obtain
the following equation of motion
\be \label{4.6}
\frac{{\rm d}\svec{S}}{{\rm d}t} =
\frac{1}{2} \svec{S} \times (\nabla \times \svec{\zeta})
- \svec{S} \frac{\partial \phi}{\partial t}
- 2  (\svec{v} \cdot \svec{S}) \nabla \phi
- \svec{S} (\svec{v} \cdot \nabla \phi)
+ \svec{v} (\svec{S} \cdot \nabla \phi)
+ o(\bar v^4).
\ee
It is found that this equation is completely the same as the equation
(9.6.5) in literature \cite{03}, it is also the same as the equation
(5.5.12) in literature \cite{04} if we neglect the influence of Thomas
precession. Follow the deduction in literature\cite{03,04}, and define
\be \label{4.7}
\svec{ S ^{~\prime}} = (1 + \phi) \svec{S}
- \frac{1}{2} \svec{v} (\svec{v} \cdot \svec{S}),
\ee
we can deduce the following equation from (\ref{4.6})
\be \label{4.8}
\frac{{\rm d} }{{\rm d} t} \svec{S^{~\prime}}
= \svec{ \Omega } \times \svec{S^{~\prime}}
\ee
with the precession angular velocity
\be \label{4.9}
\svec{\Omega}
= -\frac{1}{2} \nabla \times \zeta
-\frac{3}{2} \svec{v} \times \nabla \phi.
\ee
The first term gives the Lense-Thirring precession, and the
second term gives the geodesic precession. Applying
(\ref{4.2}) and (\ref{4.3}), we get
\be \label{4.10}
\svec{\Omega} = 3 G \svec{x} (\svec{x} \cdot \svec{J}) r^{-5}
- G \svec{J} r^{-3}
+ \frac{3 G M (\svec{x} \times \svec{v})}
{2 r^2}.
\ee
This result is completely the same as those predicted by
general relativity\cite{03,04}.\\

\section{Summary and Discussions }
\setcounter{equation}{0}

In this paper, based on the concept of the coupling between the spin
of the particle and gravitomagnetic field, the equation of motion
of the spin is obtained. In post Newtonian approximation, this equation
of motion of the spin gives out completely the same results as those
given by GR, whose calculation is  based on the equation
parallel transport. So, for the precession of orbiting gyroscope,
QGGR gives out the same theoretical prediction as that of GR. \\

Other three classical tests of gravity, including
the deflection of light by the sun, the precession of the perihelia
of the orbits of the inner planets and the time delay of radar echoes
passing the sun, are discussed in the literature \cite{21b}.
Now, we know that, for all classical tests of gravity, QGGR gives
out the same theoretical predictions as those of GR.
As we have stated before\cite{09,10}, in classical
level, QGGR returns to GR. The word "return" means that
QGGR gives out the same results for all classical phenomenon as those
of GR. Because QGGR is a perturbatively renormalizable quantum
theory of gravity and can be used to explore quantum phenomenon
of gravitational interactions, QGGR can  be regarded as a
fundamental theory that was developed from GR and
the unification of quantum theory and GR.
\\

In QGGR, gravity is treated as a kind of fundamental interactions in
Minkowski space-time. The self-consistent treatment of the classical
phenomenon of gravity in QGGR is systematically formulated in this
paper and the literature \cite{21b}. For a long time, we use the language
of space-time geometry to study classical problems of gravitational
interactions, and do not  know how to study classical problems of
gravitational interactions by using language of pure physics.
Now, in QGGR, we propose a systematic method to study classical
problems of gravitational interactions by using the language of
pure physics, which can provide us many useful information on
the nature of gravity.
\\

\end{document}